\documentclass[epj,referee]{svjour}
\usepackage{midfloat}
\usepackage{amsmath}
\usepackage{amsbsy}
\usepackage[dvips]{graphicx}

\begin{document}

\title{Laser-noise-induced correlations and anti-correlations in Electromagnetically
Induced Transparency}

\author{L. S. Cruz \inst{1}, D. Felinto\inst{1}, J. G. Aguirre G\'omez\inst{1} \thanks{Permanent
address: Departamento de F\'{\i}sica, Facultad de Ciencias F\'{\i}sicas y
Matem\'{a}ticas, Universidad de Concepci\'{o}n, Av. Esteban Iturra s/n, Barrio
Universitario Concepci\'{o}n - Chile}, M. Martinelli\inst{1}, \\ P. Valente \inst{1,2},
A. Lezama \inst{2}, and P. Nussenzveig \inst{1} \email{nussen@if.usp.br}}

\institute{Instituto de F\'{\i}sica, Universidade de S\~ao Paulo, Caixa Postal 66318,
05315-970, S\~ao Paulo, SP, Brazil \and Instituto de F\'{\i}sica, Facultad de
Ingenier\'{\i}a, Casilla de Correo 30, 11000, Montevideo, Uruguay}

\authorrunning{L. S. Cruz \emph{et al.}}
\titlerunning{Laser-noise-induced correlations and anti-correlations in EIT}

\date{\today}

\abstract{High degrees of intensity correlation between two independent lasers were
observed after propagation through a rubidium vapor cell in which they generate
Electromagnetically Induced Transparency (EIT). As the optical field intensities are
increased, the correlation changes sign (becoming anti-correlation). The experiment was
performed in a room temperature rubidium cell, using two diode lasers tuned to the
$^{85}$Rb $D_2$ line ($\lambda = 780$nm). The cross-correlation spectral function for the
pump and probe fields is numerically  obtained by modeling the temporal dynamics of both
field phases as diffusing processes. We explored the dependence of the atomic response on
the atom-field Rabi frequencies, optical detuning and Doppler width. The results show
that resonant phase-noise to amplitude-noise conversion is at the origin of the observed
signal and the change in sign for the correlation coefficient can be explained as a
consequence of the competition between EIT and Raman resonance processes.}


\maketitle

\section{Introduction}

Electromagnetically Induced Transparency (EIT) has received great attention
in recent years in connection to several interesting phenomena, such as
light storage and slow light propagation\cite{Hau99,Liu01,Bajcsy03}. The
strong interaction between light and material media in this situation
has been the source of inspiration for various proposals of applications
of EIT to the quantum manipulation of information and to transfer coherence
from light to an atomic medium\cite{Liu01,Wal03,Beausoleil04}.

The strong interaction between pump and probe fields in EIT can lead to significant
changes in the noise spectra of two independent lasers after propagation in an atomic
vapor, resulting in correlation between the fields, as was first observed in
Ref.~\cite{Garrido-Alzar03}. A recent paper~\cite{Huss04} reported an experimental
investigation of the dependence of the phase correlation between both fields generating
EIT as a function of the optical depth and transparency frequency window. The phase
modulation (PM) introduced on the pump field could be read in the probe field signal
after interaction with the atoms, for a Raman detuning within the transparency window.
Nevertheless, no distinction between positive and negative correlation was reported in
that paper. The possibility of negative correlations for large Raman detunings is
discussed in Ref.~\cite{Barberis-Blostein04}, which presents theoretical calculations for
the correlation between the pump and probe fields in EIT configuration caused by the
atomic dipole fluctuations.

In this paper, we report new measurements in which this kind of correlation
originates from two independent lasers. Both fields excite an atomic sample
forming a $\Lambda$ system, resulting in a EIT situation. We define a
normalized correlation coefficient $C$, bounded by -1 and +1, and report
measurements of $C$ as a function of intensity and analysis frequency. As
the laser intensities are increased, intensity correlations become
anti-correlations. Correlations and anti-correlations as big as 0.65 and
-0.65, respectively, were observed. The results can be explained in terms
of the conversion of Phase-Noise into Amplitude-Noise (PN-to-AN) as the
lasers interact resonantly with the atomic medium. The passage from
correlation to anti-correlation can be seen as a consequence of the
passage from EIT to a Raman resonance, both present in a 3-level atom
in the $\Lambda$ configuration.

PN-to-AN conversion in atomic vapors has been studied since 1991, when its application to
high-resolution spectroscopy was first suggested by Yabuzaki \emph{et
al.}\cite{Yabuzaki91}. This conversion relies on the characteristics of diode lasers,
which exhibit excess phase noise for usual experimental
conditions\cite{Petermann91,Zhang95} while the amplitude is generally very stable. This
excess phase noise generates amplitude noise of the polarization induced in the atomic
medium\cite{Yabuzaki91,Walser94a} and the intensity fluctuation spectrum of the
transmitted light is strongly dependent on the laser linewidth\cite{Camparo99}. Since the
fields detected after the medium are given by the input fields plus the excited
polarizations, they end up acquiring excess noise in the amplitude quadrature. The
spectral noise components that match atomic resonance frequencies present larger
amplitude oscillations. In this way, it is possible to acquire information about the
medium from the power spectrum of light after the
sample\cite{Yabuzaki91,McIntyre93,Bahoura01}. The influence of laser fluctuations on the
atomic polarization has been widely studied for two-level
systems\cite{Walser94b,Anderson90a,Anderson90b} and many models have been proposed for
treating the field phase fluctuation\cite{Ritsch90,Walser94a}. The phase diffusing model
has received more attention owing to its proximity to the diode lasers extensively used
in laboratories. In a recent experiment\cite{Martinelli04}, we performed measurements of
intensity noise spectra between the $\sigma^{+}$ and $\sigma^{-}$ components of a single,
linearly polarized, exciting field in a Rb atomic sample, as a function of the magnetic
field in a Hanle/EIT configuration.  We observed correlations as well as
anti-correlations between the different polarization components, depending on the
detuning, controlled by the magnetic field. A similar experiment was performed by
Sautenkov \emph{et al.}\cite{Sautenkov05}, using two initially phase-correlated beams in
time domain, who also explained it in terms of PN-to-AN conversion\cite{Ariunbold05}. In
Ref.~\cite{Camparo05} PN-to-AN conversion has also been identified as a source of
frequency instabilities in Rb atomic clocks, and was eliminated with a buffer gas cell
that broadens the resonances through collisions.

In the present work, the emphasis is on the PN-to-AN conversion as a source of
correlation between initially independent macroscopic fields. The paper is organized as
follows: in section 2 we describe our experimental setup and in section 3 we present a
theoretical model that includes two phase diffusing fields interacting with an atomic
system. In section 4 we present our results, beginning with the experimental correlation
spectra as a function of the analysis frequency and optical intensity. Then, in section
4b, we present the numerical results and discussion. As described below, although it is
possible to extract the basic aspects of the phenomena by modeling the atomic system as
3-level atoms at rest, agreement with experimental data is considerably improved by
integrating over the atoms' different velocity classes and including all the relevant
excited atomic levels. We also found that the optical detuning is essential for
explaining the change in sign for the correlation coefficient. For a perfectly resonant
$\Lambda$ system, the EIT process is dominant and the fields become correlated but, for
an optical detuning of the order of the excited-level decay rate, the Raman process
prevails and the fields become anti-correlated.

\section{Experimental setup}

\begin{figure}[tbp]
    \begin{center}
    \vspace{0.5cm}
        \includegraphics[width=8cm]{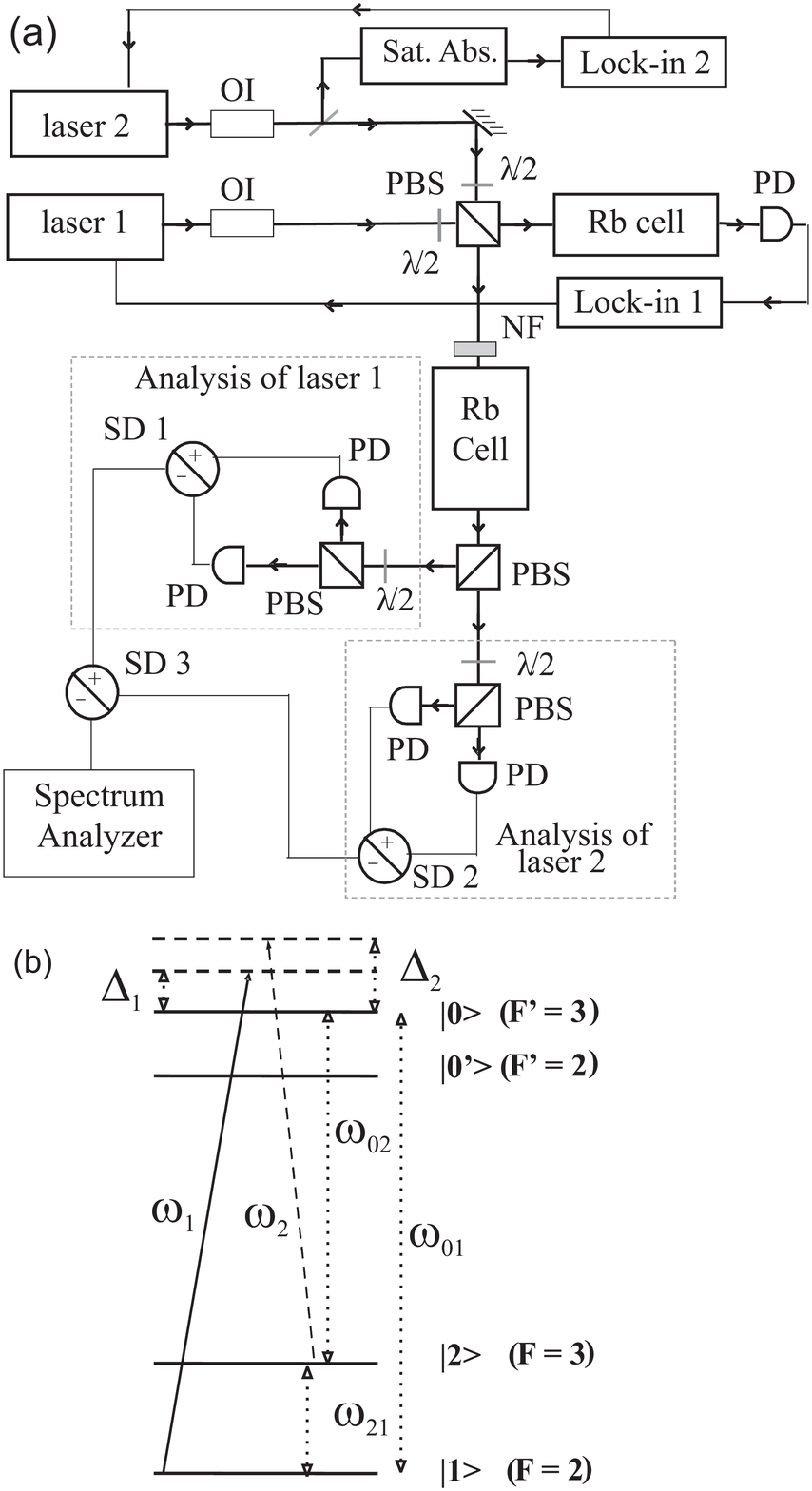}
    \end{center}
    \vspace{-0.5cm}
    \caption{\textbf{(a)} Experimental setup. The saturated absorption setup for laser 2 is
not shown. OI: optical isolator; PBS: polarizing beam splitter; PD: photo-detector; NF:
neutral filter; $\lambda/2$: half-wave plate.\textbf{(b)} Energy level representation for
our atomic model and its correspondence to the relevant Rb hyperfine states. Solid line
represents the pump field with frequency $\omega_1$; dashed line represents the probe
field with frequency $\omega_2$.}
    \label{fig:Fig1}
\end{figure}

Our experimental setup is shown in Fig.~\ref{fig:Fig1}. We employed two
external-cavity diode lasers (ECDLs) of 1~MHz linewidth and 15~mW power
after optical isolators, tuned to the Rubidium $D_{2}$ line ($\lambda=780$~nm).
The two beams had linear orthogonal polarizations and were combined in a polarizing
beam splitter (PBS). Their powers were adjusted to have equal intensities at the
vapor cell. A small portion of Laser 2 was sent to a saturated absorption setup for
fine tuning, and had its frequency locked to the cross-over peak between the
$5S_{1/2}(F=3)\rightarrow 5P_{3/2}(F^{\prime}=2)$ and $5S_{1/2}(F=3)\rightarrow
5P_{3/2}(F^{\prime}=4)$ transitions of $^{85}$Rb. The rejected output of the
polarizing beam splitter is used to observe EIT in an auxiliary vapor cell. Laser 1,
tuned to the $5S_{1/2}(F=2)\rightarrow 5P_{3/2}(F^{\prime})$ transition, is then
locked on the EIT resonance using a Lock-in amplifier (see Fig. \ref{fig:Fig1}). In
this way, we guarantee that the Raman resonance condition for EIT was always fulfilled.
Both lasers were locked only by feedback applied to their external cavity gratings.
The laser intensities were controlled by neutral density filters inserted just before
the main vapor cell. We also used a 2~mm-diameter diaphragm to spatially filter
the laser beams, ensuring a good spatial superposition and a flat, nearly top-hat,
intensity profile over the cell.

After the cell, the beams were separated at a second PBS and then analyzed at two
independent balanced-detection schemes. Photocurrents were combined in active
sum/subtraction circuits (SD), and noise was measured with a Spectrum Analyzer.
Effective bandwidth of our detection is limited by the gain of the amplifiers in the
range of 2.5 to 14 MHz. Beyond this frequency, electronic noise reduced the resolution
of our measurements. We can, therefore, measure the Sum $S_{s}(\omega)$ and Difference
$S_{d}(\omega)$ noise spectra, as well as the individual laser noise spectra
$S_{11}(\omega)$ and $S_{22}(\omega)$ by blocking a beam on each balanced detection.
It is then possible to obtain the normalized correlation coefficient defined by
\begin{equation}
C(\omega)=\frac{S_{12}(\omega)}{\sqrt{S_{11}(\omega)S_{22}(\omega)}}. \label{C}
\end{equation}
The symmetrical cross-correlation spectrum $S_{ij}(\omega)$ between the lasers $i$ and
$j$ is defined by
\begin{eqnarray}
S_{ij}(\omega) = \frac{1}{2}\langle\delta I_i(\omega)\delta I_j(\omega)^* + \delta
I_j(\omega)\delta I_i(\omega)^*\rangle.
\end{eqnarray}
Thus, $S_{12}$ can be obtained from
\begin{subequations}
\label{SumDiff}
\begin{eqnarray}
S_s = S_{11}+S_{22} + 2 S_{12} \,, \label{Sum} \\
S_d = S_{11}+S_{22} - 2 S_{12}\,. \label{Diff} \\
S_{12} = \frac{1}{4}(S_s-S_d)
\end{eqnarray}
\end{subequations}

A summary of all possibilities allowed by our setup is presented in
Table ~\ref{tab:prod}. If Laser 1 is blocked, one uses the SD2 circuit to calibrate shot
noise (with the subtraction position of SD2) or measure the total noise (with the sum
position of SD2) of Laser 2. If one blocks the Laser 2, an analogous reasoning is valid
for the SD1 circuit. The sum and difference noise spectra expressed in ~\eqref{SumDiff}
are obtained with the SD1 and SD2 switches both in the sum position and changing the SD3
switch.

\begin{table}[tbp]
\centering
\begin{tabular} {|c|c|c|c|c|}\hline
Measurement                & Beam blocked   & SD1 & SD2 & SD3\\
\hline
Total noise 1 ($S_{11}$)               & Laser 2                  & $+$ & n.i. & n.i.\\
s.n. of Laser   1                       & Laser 2                  &$-$& n.i. & n.i.\\
Total noise 2 ($S_{22}$)                 & Laser 1                    &n.i.& $+$ & n.i.\\
s.n. of Laser  2                      & Laser 1                      & n.i.& $-$ & n.i. \\
Sum ($S_{s}$)                      & none                   &  $+$ & $+$& $+$\\
Diff. ($S_{d}$)                  & none & $+$ & $+$ & $-$ \\ \hline
\end{tabular}
 \label{tab:prod}
 \vspace{0.5cm}
\caption{Summary of the different possibilities of noise measurements. s.n. means
shot-noise; n.i. stands for no influence.} \label{tab:prod}
\end{table}

\section{Theoretical Model}

\subsection{The phase diffusing field}

We developed a model based on Bloch equations. In the $D_2$ line of $^{85}Rb$ two excited
levels (out of 4 levels inside the Doppler broadened curve) can lead to EIT with the
hyperfine ground states. Inclusion of a second excited level is important to give a
better agreement with the experimental curves. The atom is excited by two classical
fields that will be considered as having constant amplitudes and independent stochastic
phase fluctuations. Our model is an adaptation of the model of Ref.\cite{Walser94a} for a
three-level atom. The four levels are represented in Fig.~\ref{fig:Fig1}, where we made a
correspondence of the hypothetical quantum states to the realistic levels of $^{85}$Rb.
The two ground states $|1\rangle$ and $|2\rangle$ correspond to $^{85}$Rb
$5S_{1/2}$($F=2$) and ($F=3$), and the excited states $|0'\rangle$ and $|0\rangle$ stand
for the $5P_{3/2}(F^{\prime}=2)$ and $(F^{\prime}=3)$ levels, respectively. We notice
that these levels form two $\Lambda$ systems for atoms of two velocity classes differing
by $kv\simeq64 $ MHz, and the EIT resonance in the room temperature vapor is built up
from nearly equal contributions of both these $\Lambda$ systems. Although all these four
levels are important for a good agreement with experimental data, for the sake of
simplicity we present an outline of calculations for a three level system (excluding the
excited $|0'\rangle $ level). The 3-level system can already reproduce many of the
experimental aspects associated to the EIT resonance \cite{Garrido-Alzar03,Martinelli04}.
Further inclusion of the fourth level is straightforward. In the following sections we
will present numerical results for both cases.

The laser fields are given by
\begin{equation}
\mathbf{E}_i(t)={\cal E}_i \exp{[i(\omega_it +\phi_i)]}\mathbf{e}_i ,
\end{equation}
where $i = 1,2$ is a label to designate lasers 1 and 2, respectively. $\mathcal{E}_i$ is
the laser's complex amplitude, $\omega_i$ its frequency and $\mathbf{e}_i$ is a unit
vector designating the field's polarization. The time evolutions of the phases
$\phi_1(t)$ and $\phi_2(t)$ are described by two independent, uncorrelated Wiener
processes\cite{Gardiner83}. This corresponds to model the lasers as phase-diffusing
fields, with Lorentzian lineshapes\cite{Walser94a}. Phase fluctuations satisfy the
relations
\begin{equation}
\langle \mathrm{d}\phi_{j}\rangle=0  , \qquad  \langle
\mathrm{d}\phi_{j}\mathrm{d}\phi_{k}\rangle=2
\sqrt{b_{j}b_{k}}\delta_{jk}\mathrm{d}t \label{mean}
\end{equation}
where $2b_{j}$ corresponds to the spectral width of laser $j$ and $\langle \cdots\rangle$
denotes stochastic average that is taken over a sufficiently long time. The $\delta_{jk}$
function accounts for the initial independence of the two lasers in our experiment, so
they have a zero degree of correlation.

For an optically thin sample, the output field can be written as
\begin{equation}
\mathbf{E}_{out}(t) = \mathbf{E}_1(t) + \mathbf{E}_2(t) +
i\frac{\beta}{2c\epsilon_0}\mathbf{P}(t) \label{E},
\end{equation}
where $\beta$ is a real constant depending on the atomic density and length of the
sample, and $\mathbf{P}$ is the complex polarization excited in the medium given by
\begin{align}
\mathbf{P}(t) & = \int_{-\infty}^{\infty} \mathrm{d}\omega_{01} g(\omega_{01})
\mathbf{p}_1(t, \omega_{01})\exp{[i(\omega_1t + \phi_1)]} + \nonumber \\ &
\int_{-\infty}^{\infty} \mathrm{d}\omega_{02}
g(\omega_{02})\mathbf{p}_2(t,\omega_{02})\exp{[i(\omega_2t + \phi_2)]} \label{P} \,.
\end{align}
In this expression, the inhomogeneous Doppler broadening is given by
$g(\omega_{0i})$, for atoms with resonance frequencies $\omega_{0i}$ in
the laboratory reference frame. $\mathbf{p}_1(t,\omega_{01})$ and
$\mathbf{p}_2(t,\omega_{02})$ are the slowly-varying atomic coherences
excited by fields 1 and 2, respectively.

The detected intensities of fields 1 and 2 are given by $I_q(t) = 2\,c\,\epsilon_0
\,|\mathbf{E}_{out}(t)\cdot\mathbf{e}_q(t)|^2$, where $q=1,2$. All power spectra can be
obtained from the expression
\begin{align}
S_{qq^{\prime}}(\omega) = \int_{-\infty}^{\infty}
[\langle I_q(t+\tau)I_{q^{\prime}}(t)\rangle
- &\langle I_q(t+\tau)\rangle \langle I_{q^{\prime}}(t)\rangle] \nonumber\\
&\times \exp{(i\omega\tau)} \mathrm{d}\tau, \label{S}
\end{align}
and we recall that the sum and difference spectra are given by Eqs.~\eqref{SumDiff}. If
we discard terms of second order in $\beta$ and terms independent of $\tau$, and use
Eqs.~\eqref{E} to \eqref{S}, we can write
\begin{align}
S_{qq^{\prime}}(\omega) = & \beta^{2}{\cal E}_{q}{\cal E}_{q^\prime}
\int_{-\infty}^{\infty}\mathrm{d}\omega_{0q}
\int_{-\infty}^{\infty}\mathrm{d}\omega_{0q^\prime}^{\prime}g_{q}(\omega_{0q})
g_{q^\prime}(\omega_{0q^\prime}) \nonumber \\ & \{ -\int_{-\infty}^{\infty}
\mathrm{d}\tau \exp{i\omega\tau} [\langle
p_q(t+\tau,\omega_{0q})p_{q^{\prime}}(t,\omega^{\prime}_{0q^\prime})\rangle \nonumber \\
& -\langle p_q(t,\omega_{0q})\rangle \langle
p_{q^{\prime}}(t,\omega^{\prime}_{0q^\prime})\rangle] + \nonumber
\\ & \int_{-\infty}^{\infty}
\mathrm{d}\tau \exp{i\omega\tau} [\langle p_q(t+\tau,\omega_{0q})p^{*}_{q^{\prime}}
(t,\omega^{\prime}_{0q^\prime}) \rangle \nonumber \\ & -\langle p_q(t,\omega_{0q})\rangle
\langle p^{*}_{q^{\prime}}(t,\omega^{\prime}_{0q^\prime})\rangle]+cc \}, \label{S2}
\end{align}

where we can see that the power spectra are obtained from the covariance matrix of the
detected intensities, which are then ultimately related to the covariance matrix for the
atomic variables $p_1=|\mathbf{p_1}|$ and $p_2=|\mathbf{p_2}|$. We now present an outline
for the calculation of the atomic covariance matrix for the case of a three level atom
excited by two phase-diffusing fields. More details can be found in Ref.
\cite{Walser94a}, especially in its Appendix B.

\subsection{Atomic polarization spectra}

The total Hamiltonian can be written as
\begin{equation}
H(t)=H_{0}+ V(t) , \label{Htotal}
\end{equation}
where $H_{0}=\hbar\omega_{01}|0\rangle\langle0|+\hbar\omega_{21}|2\rangle\langle2|$
is the free-atom Hamiltonian and
\begin{align}
V(t)= & -\hbar\Omega_{1}\exp{[i(\omega_{1}t+\phi_{1})]}|1\rangle\langle0|  \nonumber \\
& -\hbar\Omega_{2}\exp{i(\omega_{2}t+\phi_{2})}|2\rangle\langle0|,
\end{align}
is the interaction Hamiltonian, with the corresponding Rabi frequencies
for both coupling fields given by $\Omega_{j}$. Since no detailed Zeeman
structure is considered, we took both Rabi frequencies as real. We now
follow straightforward steps to write the Bloch equations, in the
Liouville form, from \eqref{Htotal} and imposing the Rotating Wave Approximation
(RWA), which results in
\begin{align}
\mathrm{d}y=exp[iN_{1}(\omega_{1}t+\phi_{1})]exp[iN_{2}(\omega_{2}t+\phi_{2})]Ax\mathrm{d}t
\nonumber\\ +y_{0}\mathrm{d}t \;.
\label{dy}
\end{align}
Here $N_{1}$ and $N_{2}$ are square diagonal matrices with only zeros and ones, $y_{0}$
is a column matrix accounting for the continuous flow of atoms through the laser beam and
$A(\Omega_1,\Omega_2, \\ \Gamma,\gamma,\Delta_1,\Delta_2,\delta_R)$ is the evolution
Bloch matrix, that is function of several physical parameters: $\Gamma$ is the total
excited state decay rate, $\gamma$ is a decay rate for the lower states coherence
associated to the finite interaction time, $\Delta_j=\omega_j - \omega_{0j}$ is the
optical detuning associated to laser $j$, and $\delta_R=\Delta_1 - \Delta_2$ is the Raman
detuning. The column matrices containing the rapid and slowly varying elements of the
atomic density matrix are $y$ and $x$, respectively. They are related by the
transformation
\begin{equation}
x=exp[-iN_{1}(\omega_{1}t+\phi_{1})]exp[-iN_{2}(\omega_{2}t+\phi_{2})]y.
\label{x}
\end{equation}
We have special interest in the $x$ matrix, because it contains the slowly varying atomic
coherences ($p_{01},p_{02},p_{12}$, and their conjugates). To proceed with the
calculations of the stochastic averages one must expand the exponential factors up to
second order in the $\mathrm{d}\phi_{j}$'s and take averages using eq.\eqref{mean},
resulting in a differential equation for $\langle x\rangle$
\begin{equation}
\mathrm{d}\langle x\rangle = [-A_{1}\langle x\rangle
+y_{0}]\mathrm{d}t
\end{equation}
with
\begin{equation}
A_{1}=iN_{1}\omega_{1} + iN_{2}\omega_{2} + b_{1}N_{1}^{2}+
b_{2}N_{2}^{2}-A \label{A1},
\end{equation}
whose steady state solution is
\begin{equation}
\langle x\rangle=A_{1}^{-1}y_{0} .
\end{equation}

However, products in the form $\langle p_q(t+\tau,\omega_{0q})p^{*}_{q^{\prime}}
(t,\omega^{\prime}_{0q^\prime})\rangle$ and $\langle p_q(t+\tau,\omega_{0q})
p_{q^{\prime}}(t,\omega^{\prime}_{0q^\prime})\rangle$ appear in equation \eqref{S2}.
To evaluate these terms it is convenient to first calculate the second order
correlation function
\begin{equation}
\langle G(t,t;\omega_{0j},\omega_{0k})\rangle=\langle
x(t,\omega_{0j})x^{\dag}(t,\omega_{0k})\rangle ,
\end{equation}
and then calculate
\begin{align}
\langle c_{2}(t,t;\omega_{0j},\omega_{0k})\rangle=\langle
G(t,t;\omega_{0j},\omega_{0k})\rangle - \nonumber \\
\langle x(t,\omega_{0j})\rangle \langle
x^{\dag}(t,\omega_{0k})\rangle,
\end{align}
where $x^{\dag}$ represents the hermitian conjugate of $x$. Finally, we use the
regression theorem to compute $\langle G(t+\tau,t;\omega_{0j},\\ \omega_{0k})\rangle$. To
obtain an equation of motion for $\langle G(t,t;\omega_{0j},\omega_{0k})\rangle$ we use
the definition \eqref{x}, differentiate the right-hand-side keeping up to second order
terms in the stochastic phases and use \eqref{dy}, resulting in
\begin{align}
&\mathrm{d}  \langle G(t,t;\omega_{0j},\omega_{0k})\rangle = \nonumber \\ &
\{-A_{1}(\omega_{0j})\langle G(t,t;\omega_{0j},\omega_{0k})
-\langle G(t,t;\omega_{0j},\omega_{0k})\rangle A_{1}^{\dag}(\omega_{0k}) \nonumber \\ &
+2b_{1}N_{1}\langle G(t,t;\omega_{0j},\omega_{0k}) \rangle N_{1} 
+2b_{2}N_{2}\langle G(t,t;\omega_{0j},\omega_{0k})\rangle N_{2} \nonumber \\ &
+y_{0}\langle x^{\dag}(t,\omega_{0j})\rangle 
+ \langle x(t,\omega_{0j})\rangle y_{0}^{\dag}\}\mathrm{d}t \label{dG}.
\end{align}

This and the use of the regression theorem allow one to get an equation of motion
for $\langle c_{2}(t,t+\tau;\omega_{0j},\omega_{0k})\rangle$
\begin{equation}
\frac{\mathrm{d}}{\mathrm{d}\tau}\langle c_{2}(t,t+\tau;\omega_{0j},\omega_{0k})\rangle
=-A_{1}\langle c_{2}(t,t+\tau;\omega_{0j},\omega_{0k})\rangle , \label{c2tau}
\end{equation}
which will be used in the calculation of the spectra. A possible way
to obtain a solution of eq. \eqref{c2tau} is to take its Laplace
transform
\begin{align}
\mathcal{G}(s;\omega_{0j},\omega_{0k})= [s+A_{1}]^{-1} \langle
c_{2}(t,t;\omega_{0j},\omega_{0k})\rangle
\end{align}
where $\mathcal{G}(s;\omega_{0j},\omega_{0k})$ is the Laplace transform of
$\langle c_{2}(t,t+\tau;\omega_{0j},\omega_{0k})\rangle$, and $\langle
c_{2}(t,t;\omega_{0j},\omega_{0k})\rangle $ can be calculated using the
steady state solution of \eqref{dG}. Since the Laplace transform is related to
the Fourier Transform by
\begin{align}
\int^{\infty}_{-\infty}f(\tau)\exp{i\omega\tau}\mathrm{d}\tau &
=\int^{\infty}_{0}f(\tau)\exp{i\omega\tau}\mathrm{d}\tau+ \nonumber
\\ & \int^{\infty}_{0}f(\tau)^{\dag}\exp{-i\omega\tau}\mathrm{d}\tau \nonumber \\ & =
\mathcal{G}(s=-i\omega) +  \mathcal{G}^{\dag}(s=i\omega),
\end{align}
the solutions $\mathcal{G}(s;\omega_{0j},\omega_{0k})$ will be used to obtain
the final results of \eqref{S2}. To include the fourth level one just adds a
new level in the Hamiltonian (eqs. \eqref{Htotal}) and repeats the calculation.

\begin{figure}[tbp]
    \begin{center}
        \vspace{2.0cm}
        \includegraphics[width=6.5cm]{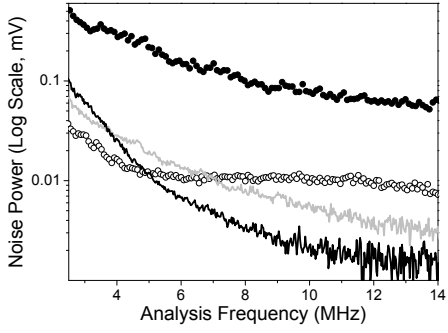}
    \end{center}
    \vspace{-2.0cm}
    \caption{Individual laser noise for both lasers after interaction with the
    atomic sample. Circles: Laser 2 (full) and Laser 1 (hollow) at high power
    ($I=118$ mW/cm$^2$); Lines: Laser 2 (black) and Laser 1 (gray) at low power
    ($I=25$ mW/cm$^2$). Resolution Bandwidth (RBW) 1.0 MHz, Video Bandwidth (VBW)
    3 kHz. Each curve is an average over 100 measurements.}
    \label{fig:Fig2}
\end{figure}

\begin{figure}[tbp]
 \vspace{-1.0cm}
    \begin{center}
        \includegraphics[width=6.5cm]{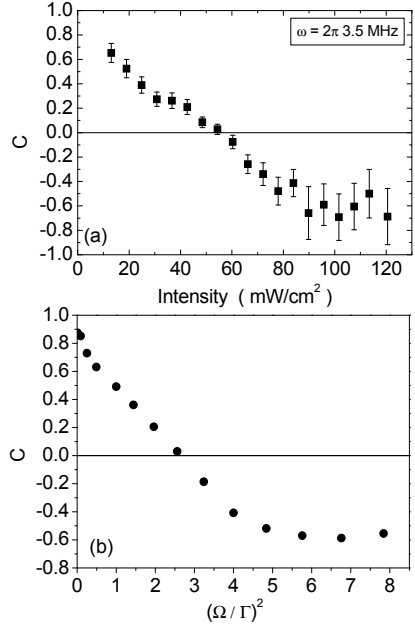}
    \end{center}
    \caption{Variation of the correlation coefficient with laser intensity
(per beam) for an analysis frequency of 3.5 MHz. \textbf{(a)} Experiment. RBW = 1 MHz and
VBW = 3 kHz. Each point is an average over $6\times 10^{4}$ measurements. \textbf{(b)}
Theory (as described in \S\ref{subsec:numerics}). $\Gamma$ is the spontaneous emission
decay rate from the excited state ($\Gamma \approx 2\pi \times 6$ MHz). Other parameters
are $\Delta_1=\Delta_2=2\pi \times 28.6$MHz, $b_1=b_2=0.08\Gamma$, $\gamma=0.02\Gamma$.}
    \label{fig:Fig3}
\end{figure}

\section{Results}
\label{sec:results}

\subsection{Experimental results}

The first step to measure the correlation between the fields transmitted by the atomic
sample is to characterize the individual noise spectrum of each field. Before
interaction, the intensity noise of the each ECDL is slightly above the standard quantum
limit (SQL). In contrast, in spite of their narrow linewidths ($\sim$ 1 MHz), both lasers
have large amounts of phase noise producing a very broad background
spectrum\cite{Zhang95}.

After interaction with the Doppler broadened atomic sample, the transmitted fields
present high degrees of intensity fluctuations. The intensity noise spectra extend to
frequencies as high as the Doppler width of the sample\cite{Yabuzaki91}. In the $D_{2}$
line of $^{85}Rb$, the energy separations of all excited levels are smaller than the
Doppler width. Thus, in a vapor cell each laser excites all the atomic transitions
allowed by dipole selection rules. The line strength of the $F \rightarrow F'-$group is
proportional to the mean value averaged by all possible dipole transitions belonging to
this group. Substituting in the measured values, the $F=3 \rightarrow F'-$group has an
effective dipole moment 1.12 times greater than the $F=2 \rightarrow F'-$group. If both
fields have equal intensities (as is the case here), the Rabi frequency associated to
Laser 2 will always be greater than the one associated to Laser 1. As a consequence, for
a sufficiently high power, the transmitted intensity fluctuation of Laser 2 will be
greater than that of Laser 1.

In Fig. \ref{fig:Fig2} we present these noise spectra for both lasers at two different
intensities measured after interaction with the atomic medium. For high power, the
PN-to-AN conversion is considerably more efficient for the laser that is locked to the
$F=3 \rightarrow F'-$group transition (laser 2), than for the laser locked to the $F=2
\rightarrow F'-$group transition. However, for a sufficiently low power, the efficiency
of the PN-to-AN conversion is very low (and comparable) for both fields, and the noise
power of Laser 2 can be lower than Laser 1 for small analysis frequencies. In this case,
we have to keep in mind that absorption also plays a role, attenuating the mean field
value and its fluctuations as well. The final result is a nonlinear character of the
PN-to-AN conversion process.

\begin{figure}[tbp]
    \begin{center}
    \vspace{3.0cm}
    \includegraphics[width=10cm]{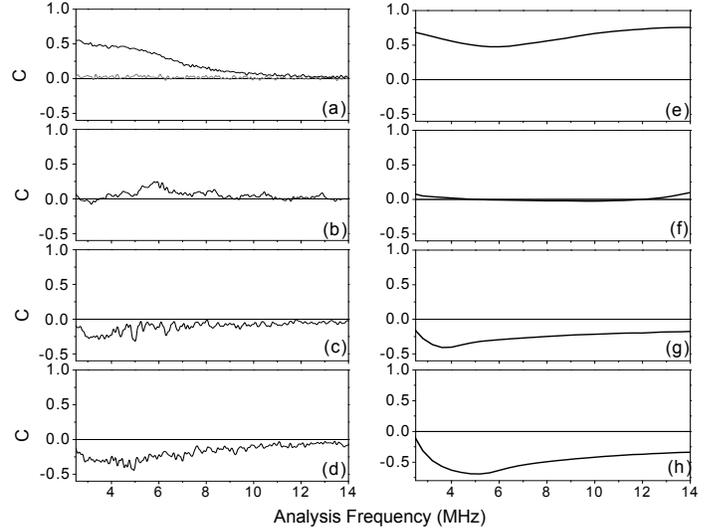}
    \vspace{-3.5cm}
    \end{center}
    \caption{\textbf{(a)-(d)} Experiment. Correlation coefficient spectra for various
    field intensities (in mW/cm$^2$). (a) 13, (b) 61, (c) 96 (d) 118. The gray curve in
    (a) gives the same measurement without the vapor cell and for the highest intensity.
    RBW = 1.0 MHz, VBW = 3 kHz. Each curve is an average over 100 measurements.
    \textbf{(e)-(h)} Theoretical result for the correlation coefficient spectra for
    various Rabi frequencies ($\Omega$): (e) $\Omega = 0.8 \Gamma$, (f) $\Omega = 1.6
    \Gamma$, (g) $\Omega = 2.0 \Gamma$ and (h) $\Omega= 2.4 \Gamma$. Other parameters are
    $\Delta_1=\Delta_2=2\pi \times 28.6$MHz, $b_1=b_2=0.08\Gamma$, $\gamma=0.02\Gamma$.}
    \label{fig:Fig4c}
\end{figure}

In Fig. \ref{fig:Fig3}(a) we show the experimental results for $C$ as the intensity is
increased, for a fixed analysis frequency ($\omega = 2\pi \times 3.5$ MHz). The first
important point to notice is the high degree of correlation between the two fields after
the sample, which can reach absolute values above 0.6. The other important feature is the
clear transition from correlation to anti-correlation as the intensity is increased,
passing through a nearly uncorrelated situation around 55~mW/cm$^2$.

We also measured the correlation spectral dependence which is shown in
Figs.~\ref{fig:Fig4c}(a)-(d) for different intensities. We observe that the transition
from correlation to anti-correlation occurs for all analysis frequencies from 2.5~MHz up
to 14~MHz. Outside this spectral range, electronic noise prevents us from measuring the
four power spectra necessary to evaluate $C$. At 14~MHz the correlation almost vanishes
for all intensities. Nearly zero correlation is observed in the range from 50 to
60~mW/cm$^2$, with small fluctuations depending on the analysis frequency. Outside this
range, the beams are clearly correlated over all the observed spectrum. Without the vapor
cell in the beam pathway (gray curve in Fig.~\ref{fig:Fig4c}(a)), the correlation goes to
zero for all analysis frequencies, as expected for two independent lasers.

The change in sign of the correlation coefficient can be understood as a consequence of
the competition between two different processes occurring in a 3-level $\Lambda$ system:
EIT and two-photon Raman transitions (both Stokes and anti-Stokes). At low intensities,
EIT generates intensity correlations between the fields, since higher intensities of one
field lead to an increase in the transparency of the medium to the second one. As the
intensity is increased, the atomic transitions are power broadened and the system becomes
saturated, so the Raman process (where one photon is absorbed from one field and emitted
in the other) dominates the atomic excitation. Since in this case the \emph{decrease} in
one field's intensity results in an \emph{increase} of the other's, it leads to an
intensity anti-correlation between them. This will be clarified below, when we
theoretically analyze the role of the optical detuning in a 3-level and in a 4-level
atom.

\subsection{Numerical results}
\label{subsec:numerics}

In order to understand the influence of the various physical parameters involved in our
experimental data, we explored our model in different ways. First, we analyzed the
situation of a 3-level atom at rest, observing the dependence of correlation with the
optical detuning, giving a physical interpretation for the change in sign of the
correlation coefficient. Next, we studied the effect of the Rabi frequency on the
correlation for the same situation, and in a second moment we performed the Doppler
integration, observing the contribution of atoms of different velocity classes. Finally,
we emphasize the contribution of the other excited atomic level presenting numerical
results for the 4-level system. These results are compared to experimental data, showing
good agreement.

The conversion of phase-noise to amplitude-noise is the main source of fluctuations
observed in this system, but it is not sufficient to explain the passage from
correlation to anti-correlation. We can have a better understanding by analyzing
what happens to the correlation coefficient in the simplified model of a 3-level
system at rest. In a $\Lambda$ configuration, the EIT resonance occurs in a
frequency window (usually) much smaller than the natural linewidth associated to
the optical transition. In other words, the two-photon Raman detuning --- expressed
as $\delta_R = \Delta_2 - \Delta_1$ following the notation of Fig. \ref{fig:Fig1}
--- should be zero and both fields must be nearly resonant with the (real) atomic
state ($\Delta_2 = \Delta_1 = 0$). If $\delta_R \neq 0$, linear absorption should occur.
On the other hand, if both fields are off resonance with the excited level and $\delta_R
= 0$, the Raman process prevails. In this case, the atom can not absorb a photon from one
of the two fields independently from the other. Only the two-photon stimulated Raman
process, in which a photon absorbed from one field is re-emitted into the other field,
can occur with high probability. We understand that the competition between these two
processes is the basis of the observed change of sign for the correlation between pump
and probe fields.

\begin{figure}[tbp]
    \begin{center}
        \includegraphics[width=6.5cm]{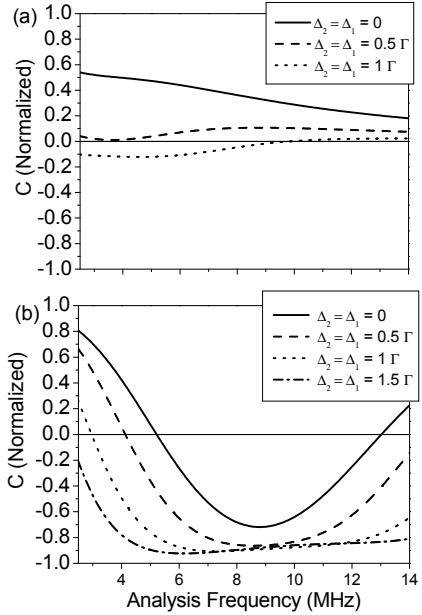}
    \end{center}
    \vspace{-0.5cm}
    \caption{Numerical calculation of the correlation coefficient $C$ for a 3-level atom
    at rest as function of analysis frequency for different optical detunings.
    Two power ranges were analyzed: (a)$\Omega_1 = 0.1\Gamma,$ (b)$\Omega_1 = \Gamma.$
    Other parameters are $\Omega_2 =1.12 \Omega_1$, $b_1=b_2=0.08\Gamma, \, \gamma=0.02\Gamma$.}
    \label{fig:Fig5}
\end{figure}

In Fig.~\ref{fig:Fig5} we show results that support our arguments. We numerically
calculated the correlation coefficient $C$ as a function of analysis frequency for a
3-level atom at rest with constant Rabi frequency $\Omega$ and different values of the
optical detuning $\Delta$. The calculations for low field intensity, presented in
Fig.~\ref{fig:Fig5}(a), give a nearly flat spectrum, with a reduction in the absolute
value of correlation for higher analysis frequencies, following the behavior expected
from the limited linewidth of the laser phase-noise. In these curves, we can see clearly
the change from correlation to anti-correlation with an increasing detuning. This effect
can be interpreted as the passage from the resonant EIT to a nonresonant Raman process,
accompanied by a change in the photon statistics. For a higher intensity --
Fig.~\ref{fig:Fig5}(b) -- the anti-correlation approaches its limit even for higher
analysis frequencies. This can be seen as a consequence of the power broadening produced
by the growth of intensity, increasing the contribution of the Raman process. Therefore,
two mechanisms are present. While detuning reduces the EIT process and the intensity
correlation, power broadening increases the Raman process and the anti-correlation.

\begin{figure}[tbp]
    \begin{center}
        \includegraphics[width=6.5cm]{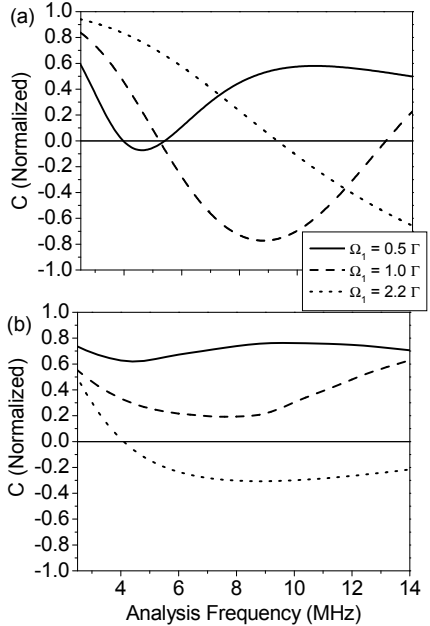}
    \end{center}
    \vspace{-0.5cm}
    \caption{Numerical results of the correlation coefficient as a function of the
analysis frequency for various Rabi frequencies ($\Omega$) in the case of a 3-level atom
at rest (a) and for a Doppler broadened ensemble (b). $\Omega_1 = 0.5 \Gamma$ (solid),
$\Omega_1 = 1.0 \Gamma$ (dashed), $\Omega_1 = 2.2 \Gamma$ (dotted).
$\Omega_2=1.12\Omega_1, \, \Gamma \approx 2\pi \times 6$ MHz for $^{85}$Rb, $b_1=b_2=
0.08 \Gamma , \, \gamma=0.02\Gamma$, $\Delta_1=\Delta_2=0$.}
    \label{fig:Fig_Fig6}
\end{figure}

A more detailed study of the effect of the field intensity can be seen in
Fig.~\ref{fig:Fig_Fig6}(a). Here we analyze an atom at rest, with zero detuning.
We can see the change from correlation to anticorrelation as a consequence of the
increase in the Raman process, together with a broadening of the shape of the curve,
demonstrated by an increase in the frequency for which the correlation changes sign.
This can also be associated with power broadening of the atomic transition.
Although we can see a few similarities with the experimental case, such as the change
of in sign of the correlation coefficient $C$ with the incident intensity, the
correlation changes rapidly with the analysis frequency, differently from what is
observed in the experiment.

A better agreement to experimental data is obtained in Fig.~\ref{fig:Fig_Fig6}(b), where
the Doppler integration was performed. In fact, it is well known that in the $\Lambda$
configuration with co-propagating fields, atoms belonging to all different velocity
classes contribute homogeneously to the signal, so if one is calculating the mean values,
the Doppler integral can be avoided. However, for a phase diffusing field the optical
detuning is a very important parameter and the Doppler width must be taken into account.
One can see intuitively that a two-level atom at rest perfectly resonant with the field
is almost insensitive to the field phase fluctuation, because it is at the maximum of the
absorption curve. In contrast, if the atom is at the maximum slope of the absorption
curve (or at the maximum of the dispersion curve), a small phase fluctuation will induce
a large intensity fluctuation in the absorption profile and, as a consequence, in the
transmitted light. This is the reason why the PN-to-AN conversion is associated to the
real part of the atomic polarization\cite{Yabuzaki91,Walser94a}. Thus, the Doppler
integral accounts for the atoms having all the possible optical detunings and the
PN-to-AN conversion in an atomic vapor is better reproduced. Furthermore, since the only
source of noise in the theory is the fluctuating phases of the incident fields, this
indicates that PN-to-AN conversion is the basic process behind our experimental
observations.

With the integration over the Doppler width, we finally observe a curve that has
a change from correlation to anti-correlation, but with a profile that changes slowly
with the analysis frequency and doesn't reach the high values of correlation calculated
for an atom at rest. Fig.~\ref{fig:Fig_Fig6}(b) still presents quantitative differences
with respect to the experimental data. For example, we see that the passage to
anti-correlation occurs for a broad range of analysis frequencies, higher than
4~MHz, while in the experiment, this passage occurs for smaller frequencies. As
seen below, the inclusion of the fourth level in the model provides better
agreement with the experimental data.

A better description of our system is obtained by including the second excited level
$|0'\rangle$ shown in Fig.~\ref{fig:Fig1} in the numerical calculation. In
Fig.~\ref{fig:Fig4c}(e)-(h) we show results for the correlation coefficient $C$ as a
function of the analysis frequency, for different values of Rabi frequencies, in the case
of the 4-level system. In this situation, each transition has a different atomic dipole
moment, so the field intensity is parameterized by a global constant $\Omega$, which is
proportional to the field amplitude and the atomic dipole moment. We chose such a range
of field intensities in order to adjust the theoretical results to the experimental
curves. We notice that in the experimental situation the lasers were not locked to a real
atomic transition corresponding to atoms at rest. Instead, we used a saturated absorption
scheme to lock laser 2 on a cross-over peak, and the other laser was locked on the EIT
resonance formed by the superposition of both. In this sense, the zero velocity atomic
class was detuned approximately 28.6 MHz above the $F'=3$ level. Since in the model the
zero energy reference for the excited state is taken on the $|0\rangle$ state, we had to
include an optical detuning in order to get a better agreement with the experimental
curves. If this optical detuning is not considered, the anti-correlation is significantly
reduced (in absolute value), but the spectral feature of $C(\omega)$ does not change
appreciably. We can observe the change from correlation to anti-correlation for
increasing Rabi frequencies, and a good qualitative agreement of the spectral plots,
especially for higher Rabi frequencies.

In order to compare these results with our experimental data, we also calculated the
variation of $C$ with the Rabi frequency for a fixed analysis frequency
($\omega=2\pi\times 3.6$ MHz). This is shown in Fig.~\ref{fig:Fig3}(b). We clearly see a
change in sign for the correlation between pump and probe fields as in the experimental
case, with a good agreement to the experimental data.

Finally, we now briefly comment the role of the laser linewidth on the sign of
$C(\omega)$. In light of the previous analysis, the correlated fields in the EIT
situation become anti-correlated if either the optical detuning is increased or the
atomic transition becomes power broadened. We checked numerically that if the laser
linewidth increases so much that it is comparable to (or higher than) the excited state
decay rate $\Gamma$, the correlation between fields tends to change sign. The physical
mechanism is totally analogous since the effect of laser broadening is to produce more
sidebands in frequencies that are not perfectly resonant with the EIT transition,
favoring the Raman process. We also confirmed that in a 3-level system this effect is
more pronounced than in a 4-level atom. In other words, for the same laser linewidth and
power, $C(\omega)$ is more negative in the case of a 3-level system than in the 4-level
situation. Consider that the carrier laser frequency is resonant with one of the excited
levels. In the 3-level atom, the laser side bands will be far off resonance with the
atomic transitions, producing pure Raman transitions, while in a 4-level atom, these side
bands approach the other excited state forming a second $\Lambda$ system (eventually
becoming resonant), so both processes tend to compensate each other.

It is important now to address some significant differences between the
experimental and theoretical results presented in Fig.~\ref{fig:Fig4c}. The main
discrepancy observed is that the theoretical curves do not show the fast decay
of the correlation as the analysis frequency is increased. Discrepancies in the
higher frequency domain can be accounted for by the amplifier gain and the
reduction of the signal-to-noise ratio of our detection. Moreover, in the
experimental curves, the Spectrum Analyzer measures the noise power centered
at a chosen frequency and averaged over a Resolution Bandwidth of 1~MHz. In the model,
the analyzer is supposed to be ideal.

Differences may also come from the simplicity of our model. In the Doppler broadened
$^{85}$Rb $D_2$ transition, the excited level is composed of 4 states, namely $F'=$ 1 --- 4,
two of which (the $F'=2$ and $F'=3$ considered in the model) contribute to the
$\Lambda$ level-schemes. However, three of the four levels contribute to the each of
the single-photon optical transitions, which give rise to the PN-to-AN process. These
are crucial for the $S_{11}(\omega)$ and $S_{22}(\omega)$ spectra -- used for determining
the correlation coefficient $C$. An important point comes from the absence of the $F'=4$ level
in calculating $S_{22}$, because this is the strongest optical (and closed) transition
and is the main responsible for the increase in the noise power seen in
Fig.~\ref{fig:Fig2} for the Laser 2. A similar reasoning is valid for the $F'=1$ level
for the case of Laser 1. Furthermore, in Fig.~\ref{fig:Fig3} we see that, in the experiment,
the laser intensity corresponding to the change in sign for the correlation coefficient is much
higher than the saturation intensity. In the theoretical case, the Rabi frequency
necessary for this change in sign is approximately twice the saturation. We remember
that the model is based on the assumption of an optically thin medium, which is not
necessarily true in the experiment. So, the real field intensity required to overcome
the EIT effect and introduce anti-correlations via Raman transitions may be considerably higher.

The way we model the laser noise may also give rise to discrepancies with the experiment.
We considered that the lasers have perfect Lorentzian lineshapes, which have slow-decay
spectral wings. A more realistic model for a diode laser, on the other hand, should
include a gaussian cutoff to the Lorentzian lineshape\cite{Dixit80,Ritsch90,Anderson90a},
so that the wings of the field spectra fall much faster than for a pure Lorentzian shape.
This comes from the fact that the amplitude and phase fluctuations are strongly coupled
in a solid state laser\cite{Henry82,Henry83}. Noise spectra, from PN-to-AN conversion,
will also depend on the shape of the phase noise spectra. This gaussian cutoff may
account for the smaller absolute values of correlation obtained in the experiment, in
comparison with the theoretical data, as well as the faster reduction of the correlation
for higher analysis frequencies. Models different from the one adopted here may give a
better agreement to the experimental data, but their implementation is a much harder task
using the present framework. Nevertheless, the simplified Lorentzian description already
gives us a good understanding of the physical processes involved, and quite good
agreement to the experimental data.

\section{Conclusions}

In summary, we have shown that the propagation of two initially independent fields generating
EIT in a vapor cell results in a high degree of correlation between the fields for a broad
range of analysis frequencies. We also observed the transition from intensity correlation to
anti-correlation as the field intensities are increased. We explain these observations in
terms of conversion of phase noise to amplitude noise by the atomic medium, and the
opposite behaviors of the EIT and Raman resonances. We develop a more detailed
calculation to support this claim. In this way, these observations reveal new basic
features of the EIT effect, and stress again how deeply EIT can affect the excitation fields.

\section*{Acknowledgments}
This project was partially supported by Funda\c{c}\~ao de Amparo \`a Pequisa do
Estado de S\~ao Paulo (FAPESP) and Conselho Nacional de Desenvolvimento Cient\'\i fico e
Tecnol\'ogico (CNPq) (Brazilian Agencies), Fondo Clemente Estable and CSIC (Uruguayan Agencies).


\end{document}